# Studying complex tourism systems:
# a novel approach based on networks derived from a time series

Rodolfo Baggio

Master in Economics and Tourism and
Dondena Center for Research on Social Dynamics,
Bocconi University, Milan, Italy
rodolfo.baggio@unibocconi.it

## Abstract

A tourism destination is a complex dynamic system. As such it requires specific methods and tools to be analyzed and understood in order to better tailor governance and policy measures for steering the destination along an evolutionary growth path. Many proposals have been put forward for the investigation of complex systems and some have been successfully applied to tourism destinations. This paper uses a recent suggestion, that of transforming a time series into a network and analyzes it with the objective of uncovering the structural and dynamic features of a tourism destination. The algorithm, called visibility graph, is simple and its implementation straightforward, yet it is able to provide a number of interesting insights. An example is worked out using data from two destinations: Italy as a country and the island of Elba, one of its most known areas.

## 1. Introduction

A tourism destination is a complex and complicated ensemble of diverse components of interrelated economic, social and environmental factors, all deeply connected among themselves. It has been recognized to be a changing dynamic system, in which sparking events, both internal or external, natural or human, can challenge existing configurations, normal operations or even the very existence of the system and can dislodge it from an equilibrium state towards different and erratic evolutionary paths. All this with a very little predictability, which makes problematic the governance of the system and the design of strategies for improving the overall effectiveness and efficiency of both the whole and its components (Farrell & Twining-Ward, 2004; Faulkner & Russell, 2001).

As a research topic, tourism is well suited for interdisciplinary research (Przeclawski, 1993). Approaches and methods originating from an assorted range of disciplines, such as economics, geography, sociology, management, have been used to understand the nature and behavior of the tourism phenomenon which is characterized by poorly defined boundaries and comprises a multiplicity of organizations offering heterogeneous products and services (Mazanec & Strasser, 2007).

These products may be considered to be collections of components (such as accommodation, transport, attractions, hospitality etc.), where the relationships between the different elements are difficult to define and analyze in aggregate form due to the variability in which different customers arrange them throughout their trip. A number of models, ideas and methods have been used to study tourism systems (Cooper et al., 2005), but many often raised problems in the capability of fully describing the complex and dynamic socio-economic environments of tourism. In particular, they have had little success in providing satisfactory insights into the possible development paths of such systems (Farrell & Twining-Ward, 2004; McKercher, 1999).

One useful approach to the study of the tourism phenomena is to focus on tourism destinations. These are the geographic locations where tourists spend most of their time when travelling. A destination contains "a critical mass of development that satisfies traveler objectives" (Gunn, 1997: 27), and thus offers a tourist the opportunity of taking advantage of a variety of attractions and services. Many scholars consider it a fundamental unit of analysis for understanding of the whole tourism phenomenon, even if difficult to define precisely and problematic as a concept (Framke, 2002).

A destination has the properties of a system: an organized assembly of elements or parts (components) connected to each other with some defined relationship, and having the general objective of accomplishing a set of specific functions, or achieving particular goals (Ashby, 1956; Carlsen, 1999). The systemic approach provides a broad framework that allows different perspectives to be used flexibly in the study of tourism, rather than assuming rigid predetermined views. It enables an understanding of the broad issues which affect tourism and takes into consideration the relationships between its different components (Page & Connell, 2006).

Identification of tourism destinations as systems is a useful analytical approach, but stimulates further questions on what type of system it is, what are its components and how their interactions affect the overall dynamics of the system. In a pioneering work, Faulkner and Valerio (Faulkner & Valerio, 1995), considering the deficiencies and the unreliability of many prediction and forecasting methods for tourism, called for the use of alternative ways to explain tourism dynamics, and proposed the adoption of a chaos and complexity framework. Since then a growing strand of literature has recognized the complexity characteristics of tourism systems noting the non-linearity of the relationships that connect the different companies and organizations, and the response of the various stakeholders to inputs that may come from the external environment or from what happens inside the destination (Baggio, 2008; Haugland et al., 2011). Obviously, not all destination systems share exactly the same characteristics and behaviors, and diagnosing the extent to which a destination may be considered a stable, a complex or even a chaotic system can be of great interest not only from a theoretical point of view, but also because it may provide crucial insights into the possibility of governing and steering the destination towards a desired evolutionary path (Baggio & Sainaghi, 2011; Baggio et al., 2010a).

This diagnosis can be done by employing different methods with different degrees of sophistication and intricacy. One recent proposal, however, seems to be relatively simple and straightforward and, even with limitations, able to provide at least a first answer to the problem of assessing the 'complexity' of a system (Lacasa et al., 2008). It relies on an observable series of data taken as representative of the dynamic behavior of a system and uses a mapping of this series into a network. In this way the powerful methods of network science can be used for the investigation.

Aim of this paper is to present this type of analysis and to provide methodological guidance. The results allow us to uncover the main characteristics of the system and to highlight a new way to understand the dynamics of tourism development. The rest of this paper is organized as follows. Next section briefly sketches the different possibilities for approaching the analysis of a tourism destination from a complex system science perspective. The method proposed is then explained in detail. The subsequent section presents the investigation of two examples, one at a country level (Italy), one at a local level (Elba island, Italy). The final section contains closing considerations and addresses limitations and possible future works.

## 2. The study of tourism destinations as complex systems

The application of different complexity science methods, well known in physics, mathematics sociology and economics, but not widely used in the tourism literature, has provided already a good array of insights into the structure and the dynamic behavior of a tourism destination. The general complexity characteristics of a tourism system have been explored by using non-linear time series analysis techniques, agent based numerical simulations and by applying complex network analysis methods (Baggio, 2011a; Baggio & Baggio, 2013; Baggio & Sainaghi, 2011; Baggio et al., 2010b; Cole, 2009; Johnson & Sieber, 2010; Scott et al., 2008, 2011).

The network science approach has uncovered important outcomes concerning destinations' structures, the functioning of collaborative and cooperative groups, the diffusion of information or knowledge across the system or the relationships between the physical and the virtual components of a destination. Additionally, the network approach has been extended to implement simulation models with which different scenarios can be obtained in order to explore the possible effects of different managerial or governance activities. This provides all those interested in the life of a tourism destination with powerful tools to inform their policy or management strategies. The network perspective can offer a number of useful outcomes for tourism studies, but has also shown some limitations mainly due to the difficulty of collecting the data needed to perform a full analysis (Scott et al., 2011).

Other techniques, successfully used in many different disciplines use non-linear analysis methods applied to observational time series (Kantz & Schreiber, 1997; Sprott, 2003). Popular methods employed in a variety of applications include: Lyapunov exponents, fractal dimensions, symbolic discretization, and measures of complexity such as entropies or quantities derived from them. All these techniques have in common that they measure certain dynamically invariant properties of the system under study based on temporally spaced realizations of the development paths. However, their application requires employing sophisticated techniques that rely, in many cases, on a good and deep experience and knowledge of the researchers. Moreover, all these methods require, for their best working, large amounts of data that are not very common in the tourism field. Even if some of these techniques have been successfully applied to the study of a tourism destination, despite the existence of reasonably 'usable' software tools, their usage and the interpretation of the results rests a task which can be difficult for many, especially practitioners (Baggio & Sainaghi, 2011).

Recently, however, new methods have been proposed that allow to derive general characteristics of a complex system by using a time series of observations and transforming it into a network. The idea is that it is possible to consider a time series just as a set of numeric values and play a simple game of transforming it into a different mathematical object. Then we can check what properties of the original set are conserved, what are transformed, or what can be inferred about one of the representations by examining the other. It turns out that a number of interesting insights can be derived by using this method and that this mathematical game has various unexpected practical applications opening the possibility of analyzing a time series (i.e. the outcome of a dynamical process) from an alternative perspective. Finally, since the derived representation belongs to a mature and rigorous field - network science - the information encoded in such a representation can be effectively processed and interpreted (Nuñez et al., 2012; Strozzi et al., 2009).

In this line of research different techniques have been proposed, based on concepts such as correlations, phase-space reconstructions, recurrence analysis, transition probabilities ((an extensive list can be found in Donner et al., 2010 and references therein). All these have shown that different features of a time series are mapped onto networks with distinct topological properties, thus suggesting the possibility to distinguish the properties of time series, and ultimately of the system from which they originate, using network measures (Campanharo et al., 2011; Donner et al., 2010; Yang & Yang, 2008).

Probably the simplest method, conceptually and computationally, is the one proposed by Lacasa et al. (Lacasa et al., 2008; Nuñez et al., 2012): the visibility algorithm. By using this technique it has been show that a time series structure is inherited in the associated graph, such that periodic, random, and fractal series map into networks with different topologies (random exponential or scale-free).

A visibility graph algorithm thus allows applying methods of complex network analysis for characterizing the system in a straightforward way. In the transformation, some information regarding the time series is inevitably lost due to the fact that the network structure is completely determined in the (binary) adjacency matrix, while two different series with the same periodic succession of values would have the same visibility graph, although being quantitatively different. However, the simplicity of the algorithm and its fast implementation make it a good candidate for an initial scrutiny. Moreover a visibility graph remains invariant under several transformation of the time series data such as translation, vertical rescaling, or addition of a linear trend.

So far, a number of studies have been published in fields of stock market indices, exchange rates, macroeconomic indices, human behaviors, neurology, occurrence of hurricanes, or dissipation rates in turbulent systems (Nuñez et al., 2012).

## 3. Materials and methods

The destination used as examples are Italy and the Italian island of Elba. Using a country and one of its most representative part will also allow to highlight possible similarities or differences between a system and one of its subsystems. From a 'tourism' perspective both are interesting subjects. Italy is one of the most important tourism destinations in the World. According to the rankings published by the UN World Tourism Organization (UNWTO, 2011) Italy is at the third place in Europe and fifth in the World. In 2011 roughly 104 million tourists have spent some 390 million nights in the Italian accommodation establishments. About 46% of them are international visitors. Tourism is a quite important contributor to the country's economy and accounts for about 8.5% of the GDP and occupies 9.5% of the employment (total contribution). Elba island is a typical summer destination whose economic activities are prevalently bound to tourism. It accounts (in 2011) for about 500 thousand arrivals and 2.8 million overnight stays, 32% of the tourists are international visitors.

The series used in the analysis are the monthly overnight stays series. For Italy the series spans the period 1987-2011, for Elba 1954-2011. All data come from the official Italian statistical bureau ISTAT (www.istat.it) and from the statistical office of the Livorno province where Elba is located (www.provincia.livorno.it). The difference in length between the two series (300 and 696 points) also allows to show the flexibility of the method and its relative insensibility to the amount of data used. As customarily done, the two series have been detrended.

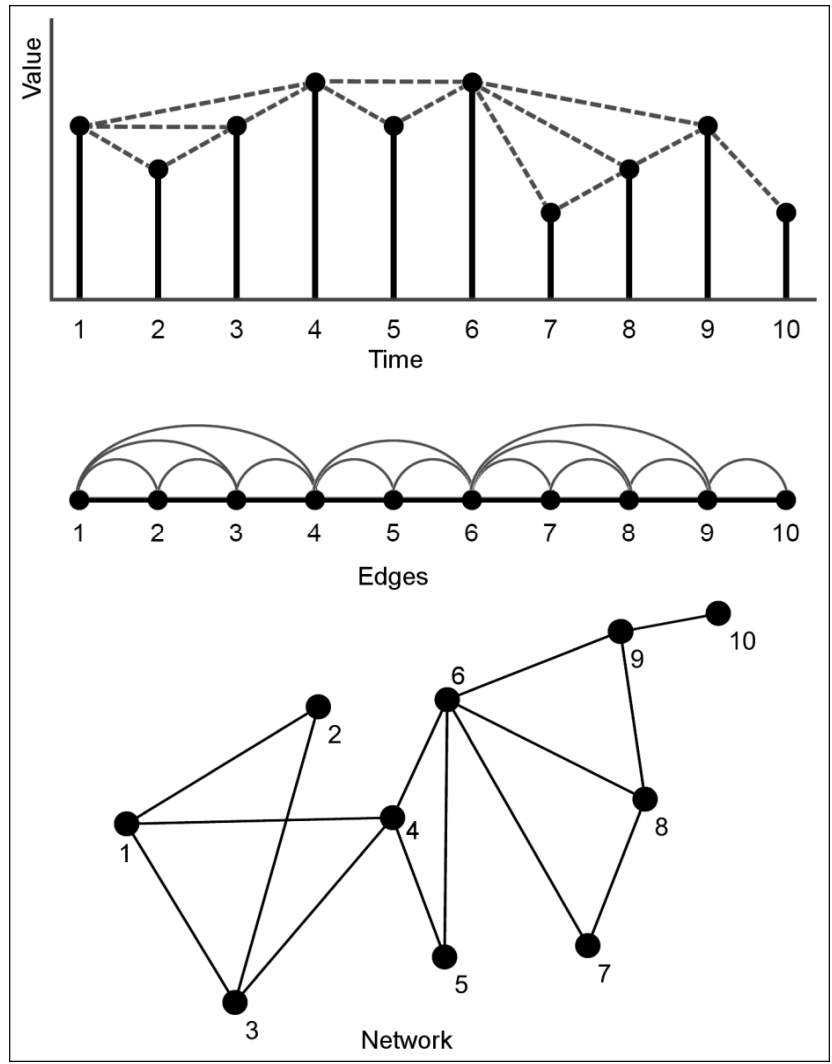

Figure 1: An example of visibility graph derived from a time series. The figure shows the original time series, the edges calculated according to the visibility condition, and the resulting network

The algorithm used for mapping the time series into a network is called visibility graph. It can be described as follows. Let us consider a time series $Y(t) = [y_1, y_2 \ldots y_n]$ of length N. Each data point $y_n$ in the series can be regarded as a vertex in the associated network and an edge can be drawn connecting two vertices if the two corresponding data points can 'see' each other in the vertical bar chart of the time series. In other words two data points are connected when there is a there is a straight 'visibility line' that joins the points without crossing any other intermediate data bar (see Figure 1 for an example).

Formally, two data values $y_a$ (at time $t_a$) and $y_b$ (at time $t_b$) are connected if, for any other value ($y_c$, $t_c$) existing between the two (i.e.: $t_a < t_c < t_b$), the following condition is satisfied:

$$y_c < y_a + (y_b - y_a)\frac{t_c - t_a}{t_b - t_a}.$$

The visibility graph algorithm is simple to program and runs relatively fast even for large datasets. The resulting network is then analyzed using standard techniques that consist of calculating the relevant metrics. Many of these quantities have been proposed in the last years and the literature contains a wealth of possible ways for assessing many of the structural and dynamic characteristics of the network both at a global and local level (for an extensive list see da Fontoura Costa et al., 2007; Newman, 2010). The next section provides a guided tour for the analysis of the destination considered.

## 4. Results and discussion

The two networks obtained are shown in Figure 2. By construction the networks contain a single component (i.e. no disconnected nodes exist). The similarity in the topologies of the networks is rather evident. This is a first, visual, confirmation of the self-similarity characteristics of the complex Italian tourism system.

In the rest of this section, loosely following similar analyses conducted in other cases (Chao & Jin-Li, 2012; Wang et al., 2012), the main characteristics of the networks and their interpretations are discussed.

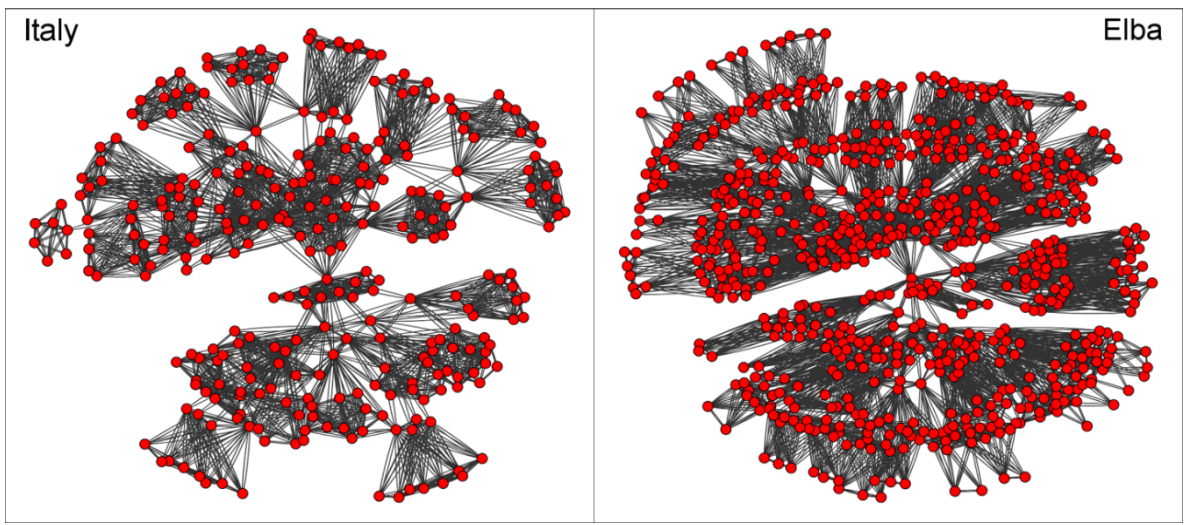

Figure 2: The networks for Italy and Elba obtained by running the visibility graph algorithm

## 4.1 Degrees distributions

The degree k of a node is the number of connections the node has in the network and measures how large is its direct influence on others. The statistical distribution of the degrees is an important parameter for a network and characterizes its nature. Many complex systems exhibit a peculiar degree distribution which follows a power law $N(k) \sim k^{-\gamma}$. That is to say that a few nodes (hubs) have a large number of connections while the vast majority has a limited number of links. In our case, a time value corresponding to a node with very large degree manifests a sharp and sudden rise or peak in touristic activities. The two degree distributions (cumulative) are shown in Figure 3a. The largest part of the curves are compatible with a power law distribution, The exponents are: $\gamma(Italy) = 2.59 \pm 0.67$ and $\gamma(Elba) = 2.54 \pm 0.73$. Here, again, we note the striking similarity of the two topologies.

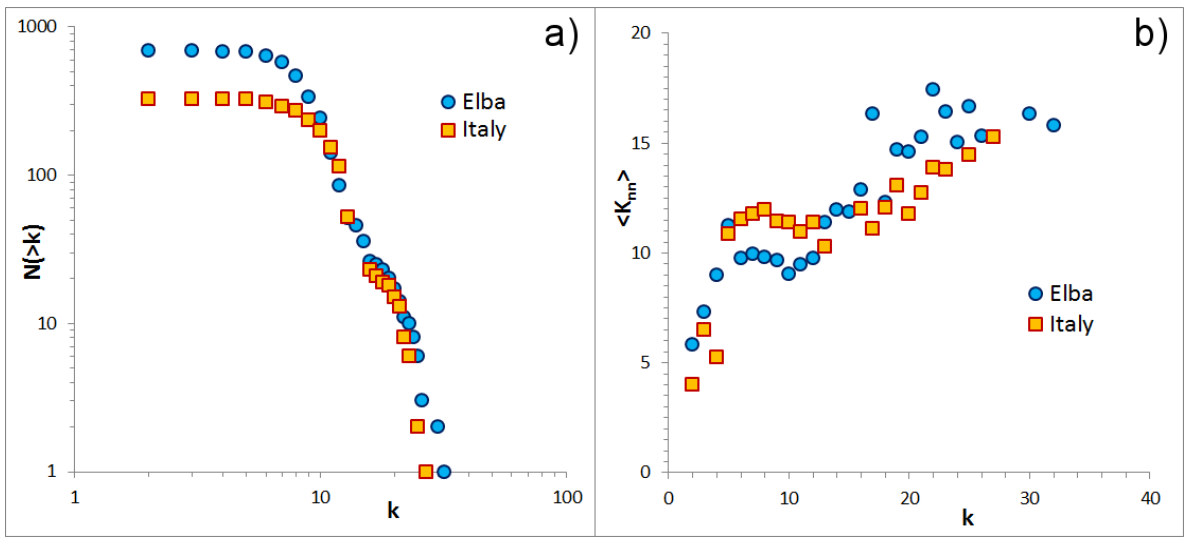

Figure 3: Cumulative degree distributions (panel a) and nearest neighbors average connectivity distributions (panel b) for both networks.

## 4.2 Average neighbor connectivity and assortativity

The form of the degree distribution *N(k)* has a direct influence on the properties of a network and accounts for its basic topology. However, it cannot convey all the information on the network structure. In fact, two networks can have similar distributions yet exhibit different static or dynamic characteristics that are, generally, determined by the presence of correlations between the degrees (Gallos et al., 2008). This structure can be captured by the probability that two nodes with different degrees connected to each other. Two quantities can provide this information: the distribution of the average degree of nearest neighbors $\langle K_{nn}\rangle$ and the Pearson correlation coefficient $r$ between the nodal degrees.

The relationship plays an important role in determining the unfolding of a propagation process (perturbations, information or influence diffusion) on the network. It is reasonable to assume that if a perturbation starts from a node (and highly connected nodes are powerful amplifiers) it can affect with a certain probability its first, second, and sometimes even more distant neighbors in the corresponding network. Moreover, the resilience of a network, that is its capacity to withstand external or internal shocks without being disrupted but recovering in a reasonable period of time, is very sensitive to degree correlations (Newman, 2002).

The Pearson correlation coefficient r accounts for the attraction or repulsion tendency between similar nodes. The metric is called assortativity in network science and, in the case of a social network, can be seen as a possible expression of the attraction existing between individuals sharing similar characteristics. As a matter of fact, many social networks show a positive assortativity, while generally a negative correlation is typical of technological or artificial networks. Concerning resilience, numerical simulations have shown that a positive assortativity imply robustness against targeting high degree nodes through redundancy, since these hubs tend to be clustered forming cohesive groups. The more assortative a network is, the higher its resilience (Serrano et al., 2007).

Figure 3b shows a clear positive relationship between $\langle K_{nn}\rangle$ and $k$ for both the destination networks examined. This is further confirmed by the positivity of the assortative coefficient for both systems: *r(Italy) = 0.138* and *r(Elba) = 0.316*.

## 4.3 Clustering coefficient

The clustering coefficient *C* measures the concentration of connections of a node's neighbors. It provides a measure of the heterogeneity of local density of links and quantifies how well connected are the neighbors of a vertex. The metric can provide an indication of the extent to which the tourism organizations work together collaborating or cooperating, i.e.: forming cohesive communities inside the destination. More importantly, the clustering coefficient can be used to uncover the hierarchical organization of the networked system. Ravasz and Barabási (2003) have shown that the relationship between the average clustering coefficient and the degree of the nodes signals a hierarchical structure when it follows a power-law functional form: $C_{ave}(k) \sim k^{-\alpha}$. As Figure 4a shows, this is valid for the main part of the distributions calculated for the destinations under study, and the slope of the curves are quite similar. In particular the values for the exponents are: *α(Italy) = 1.28±0.12*, and *α(Elba) = 1.26±0.30*.

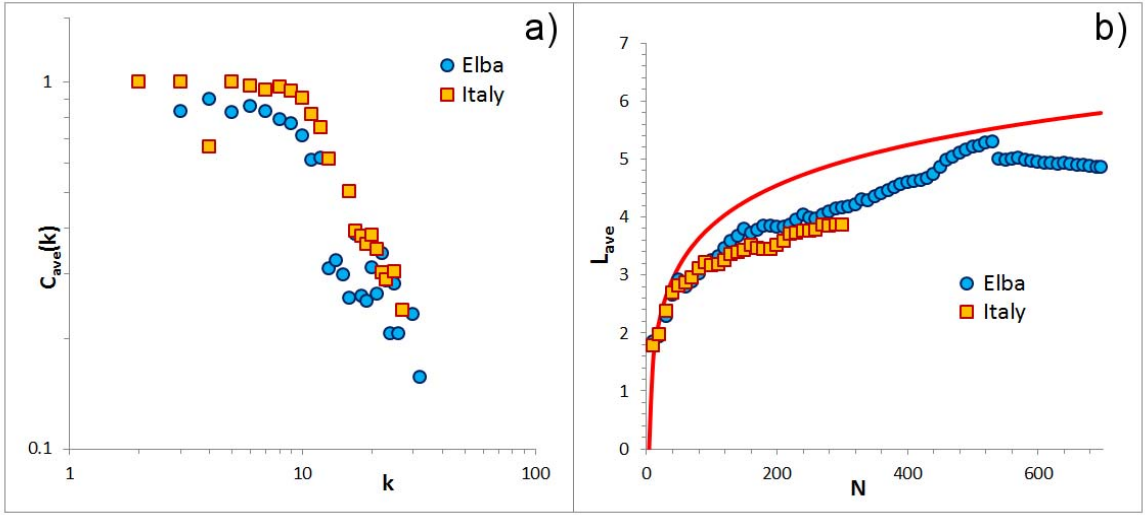

Figure 4: Average clustering coefficient as function of degree (panel a) and average path length as function of the number of nodes (panel b)

### 4.4 Average path length and small-world behavior

The average path length is the mean value of the distance (number of links) between any pair of nodes. As shown by the seminal work of Watts and Strogatz (1998), a network can exhibit a small-world behavior which is characterized by a low average path length and a high clustering coefficient, differently from what happens in a network where links are distributed randomly. Small-world is a characterizing feature of many social networks and accounts for some of the behaviors of people or groups that tend, in a social setting, to be more closely connected, mainly when displaying similarity in some of their traits.

A simple way of assessing this feature is to recall that the average path length increases logarithmically (or more slowly) with the number of nodes $N$: $L_{ave}(N) \sim ln(N)$. Figure 4b shows clearly that this is the case for our destinations, therefore the visibility graphs of both networks are small-worlds.

### 4.5 A summary

In conclusion it is possible to summarize the outcomes of the study as follows.

Both destinations exhibit the characteristics of a complex networked system and a good similarity in their topologies. This resemblance reinforces the idea of a tourism destination as a self-similar system (Elba is a subsystem of Italy and shows a comparable topology). A direct conclusion is that these are systems whose behavior is difficult to be predicted (or that the predictability window is small), and will show a good resilience in case of unforeseen events (shocks).

The destinations also exhibit a hierarchical structure which testifies the existence of an emergent self-organization behavior. One immediate consequence of this fact is that policy or governance measures that do not take into account the autonomous organization of the system are destined to have little impact (see also Baggio, 2011b).

Finally, the small-world characteristics of the networks show that the networks are relatively compact and clustered into small cohesive groups, similarly to many other social and economic networks. This feature has, among others, the consequence of easing diffusion processes that may occur on the network. In other words, once chosen the starting points, information or opinions could be transferred to a large proportion of the actors efficiently in relatively short times. Furthermore, in a small-world network it is easier to have a convergence of opinions with respect to networks exhibiting purely random distribution of connections (Wang & Chen, 2002). Last but not least, the combined effect of small-world behavior coupled with the substantial heterogeneity of the network topology has an important effect in sustaining cooperative attitudes (when they exist) among the network's actors (Santos et al., 2005).

## 5. Concluding remarks

Tourism destinations are complex adaptive systems and their complexity is a crucial characteristic which affects a number of properties of the system as well as its dynamic behavior.

Assessing the complexity of such systems has important implications both from a theoretical and a practical point of view. Different methods exist for performing a diagnosis, mostly based on non-linear analysis of series of values that represent in some way the outcomes of the behavioral conditions of the object of study or by collecting the appropriate data needed to build a complete network. Both possibilities, however, raise some issues for their inherent difficulty or for the problems met when collecting the data needed. Here a novel and relatively simple approach has been presented which uses a mapping of a time series into a network therefore allowing network science techniques to be applied. The results presented on the study of the two destinations are all well in line with those obtained elsewhere by employing non-linear and network analysis methods (Baggio, 2008; Baggio & Sainaghi, 2011; Baggio et al., 2010a, 2010b).

Other variations of the algorithm have been proposed that may highlight different possible features and make the analysis more complete. These will be object of the future efforts of this line of research.

In the opinion of the author, the main limitation in the method presented here is of conceptual nature. The hypotheses made are that the main structural and dynamic characteristics of a complex system can be rendered through a series of observations (time series) and that the transformation of the time series into a network does not lose too much information thus allowing to preserve at least the key traits. Although reasonable and verified in a number of cases, these assumptions will need better and more extensive investigations before being fully accepted. For the time being, however, the ansatz seems to work well.

**Rodolfo Baggio** has a degree in Physics and a PhD in Tourism Management. He is professor at the Master in Economics and Tourism and Research Fellow at the Dondena Center for Research on Social Dynamics at Bocconi University, Milan, Italy. He actively researches on the use of information and communication technology in tourism and on the applications of quantitative complex network analysis methods to the study of tourism destinations.